\documentclass[12pt]{article}

\usepackage{graphicx,amssymb,array}

\newcommand{\cs}{cross-section}
\newcommand{\css}{cross-sections}
\newcommand{\nN}{neutrino-nucleon}

\newcommand{\sfts}{structure functions}
\newcommand{\nuN}{$\nu N$}
\newcommand{\n}{neutrino}
\newcommand{\ns}{neutrinos}

\newcommand{\SFs}{\emph{SFs}}

\newcommand{\DL}{\emph{DL}}
\newcommand{\CR}{\emph{CR}}
\newcommand{\CRs}{\emph{CRs}}
\newcommand{\QCD}{\emph{QCD}}
\newcommand{\pQCD}{\emph{pQCD}}
\newcommand{\HE}{\emph{HE}}
\newcommand{\UHE}{\emph{UHE}}
\newcommand{\HENA}{\emph{HENA}}

\begin{document}
\title{Hard Pomeron Enhanced Cascade Production and
Flux Shadowing in High-Energy Neutrino Astrophysics}

\author{A. Z. Gazizov\thanks{gazizov@dragon.bas-net.by} \ and      
S. I. Yanush\thanks{yanush@dragon.bas-net.by}\\ \\                 
B. I. Stepanov Institute of Physics \\of the National Academy of   
Sciences of Belarus,\\ F. Skariny Ave.\ 68, 220072 Minsk, Belarus} 

\date{}        
\maketitle     


\begin{abstract}
Various implications of new, non-perturbative pomeron inspired
enhancement of small-$x$ \nN\ \sfts\ for high-energy \n\
astrophysics are discussed. At $x \gtrsim 10^{-5}$ these functions
are given by perturbative \emph{QCD}, while at lower $x$ they are
determined by a specific generalization of $F_2^{ep}(x,Q^2)$
description, proposed by A. Donnachie and P. V. Landshoff (their
two-component model comprises \emph{hard} and \emph{soft}
pomerons), to \nuN-scattering case.

We found that i) such enhancement causes the most rapid growth of
$\nu N$-\css\ at high energies, ii) that pomeron effects may be
perceptible in the rates of \n\ induced events in future giant
detectors and iii) that the rate of high-energy \n\ flux evolution
(due to absorption ($CC+NC$) and regeneration ($NC$)) on its pass
through a large column depth of matter may be subjected to
additional influence of hard pomeron. Solving transport equations
for the initially power-law decreasing $\nu$-spectra, we have
evaluated shadow factors for several column depths and spectrum
indices. The results are compared with analogous calculations,
performed within a trivial small-$x$ extrapolation of \sfts. Hard
pomeron enhanced high-energy shadow factors are found to be many
orders of magnitude lower than those obtained within ordinary
perturbative \QCD.
\end{abstract}


\section{Introduction} \label{introduction}%
As a rule, accelerator experiments provide data of very high
quality. But such measurements seem to be hardly possible at
energies, $E \gtrsim 1 \times 10^{19}$~eV, even in the very
distant future.

On the other hand, \emph{non-accelerator}, observational physics
gives a good chance to get data at ultrahigh energies
\cite{BBGGP}. But the accuracy of data in this case is worse  due
to scarcer statistics and lower resolution of detectors. It is
remarkable, that non-accelerator \emph{High-Energy (HE)} physics
combines into one the 'telescope' and 'microscope' physics.
Really, the detected \emph{UltraHigh-Energy (UHE)} particles are
produced in perhaps most distant, extragalactic cosmic sources.
Hence, it deals with extremely large, cosmological, distances. On
the other hand, interactions of such particles with matter occur
at least distances.

Realization of \UHE\ astrophysics is a complex problem. It poses
many hard questions, such as:

\vspace{1mm}%
1. what types of ultrahigh-energy particles may be
observed?

\vspace{1mm}%
2. where and how ultrahigh-energy particles are produced?

\vspace{1mm}%
3. what detectors are to be used?

\vspace{1mm}%
4. what is the expected rate of events in a detector?

\vspace{1mm}%
5. does this rate exceed the background?

\medskip
Let us try to answer these questions in brief.
\subsection{Particles}
Through the centuries people used only star light, viz.\ optic
photons, for the sky observation. Today actually the whole range
of e.m.\ radiation, up to TeV photons \cite{KASKADE}, is involved
in measurements.

Another well-known example of high-energy cosmic radiation is
provided by Cosmic Rays \emph{(CRs)}. These are mostly protons
($H$), $He$, $C$, $N$ and up to $Fe$ ions, which continuously
bombard the atmosphere. Energy spectrum of \CRs\ extends from
$1$~GeV up to perhaps $1 \times 10^{21}$~eV; the highest energy
event with $E \simeq 3\times 10^{20}$~eV has been detected by the
\emph{AGASA} array \cite{AGASA}. The nature of incident particles
in such events as well as where and how they are produced is
unknown. Moreover, the average path length of $E \gtrsim 6 \times
10^{19}$~eV protons in the cosmic microwave background radiation
(with temperature $T=2.73K$) is essentially limited by $ p +
\gamma \rightarrow \pi^+ + n$ interactions \cite{GZK} (\emph{GZK}
cut-off), so that their presence in \CRs\ is a problem.

Besides traditional photons and \CRs, gravitation waves and \ns\
are also regarded as possible instruments for \HE\ astrophysics.
In this paper we concentrate on \ns, in particular, on \UHE\ \ns\
with $E_\nu \gtrsim 1\times 10^{-15}$~eV. Being neutral, they do
not deflect in magnetic fields, hence their arrival directions
shoot back to the production site. Small \css\ allow these \ns\ to
travel extremely large distances without absorption. But their
production in cosmic sources is difficult (i.e.\ sources are rare)
and these sources are distant. It makes the expected flux to be
small. Moreover, detection of such \ns\ is complicated by the same
small \css. To compensate for these negative factors, one should
construct gigantic detectors.
\subsection{Production} There are many theoretical models
predicting large luminosity of cosmic sources in high-energy
particles, including \ns\ (see e.g.\ \cite{BBGGP}). For example,
such fluxes may be produced in supernova explosions, in active
galactic nuclei, during gamma-ray bursts etc. Energy outburst in
these sources, including in the form of \HE\ particles, may be
really high, but the expected flux at the Earth remains low due to
large distance.

Basically, there are two models of \HE\ particle production. First
assumes proton acceleration up to very high energies on cosmic
source shocks (usually gas accretion onto a massive black hole,
presumably residing in center of an active galaxy, is suggested),
sometimes in relativistic jets pointing to the Earth. Then pions
and kaons appear as a result of $pp$- and/or $p\gamma$-collisions
in a source. These mesons decay, giving birth to \HE\ \ns. The
described scenario is called \emph{down-top} mechanism.
Unfortunately, it is difficult to obtain proton energies higher
than $E \sim 10^{17}$~eV in this mechanism; one needs both large
shock radius and high magnetic field.

Another possibility is proposed by the \emph{top-down} mechanism
(for review and discussion of upper limits to the expected
$\nu$-fluxes see e.g.\ Ref.~\cite{NPQSGAZ} and references
therein). It is based on suggestion that some yet unknown,
extremely massive (with $m_X \sim 10^{14} \div 10^{16}$~GeV)
long-living gauge $X$-bosons decay to the ordinary particles.
These bosons allegedly originate from topological defects, which
in their turn have been produced during a hypothetical phase
transition in the early Universe. A big fraction of $X$-boson's
energy goes to \HE\ particles (\ns). These attractive models
involve new physics beyond Standard Model and seem a bit
speculative. An experimental verification is needed.
\subsection{Detectors} \HE\ \n\ detectors should be very
large and well shielded from background radiations. The are to
have either $S\gtrsim 1$~km$^2$, especially if secondary muons
(say from $\nu_\mu + N \rightarrow \mu^- + X$) are detected,
and/or gigantic mass, $M \gtrsim 1\times 10^9$~ton, in the case of
registration using nuclear-electromagnetic cascades (from hadron
state $X$ in the $\nu_\mu$ case or from $\nu_e + N \rightarrow
X$). These requirements may be met:
\begin{itemize}
  \item by search for Cerenkov radiation in deep underwater
  detectors with desirable volume under control $V \sim 1$~km$^3$
  (mass $M \sim 10^9$~ton);
  \item by registration of air nitrogen fluorescence, induced by
  Extended Air Showers; an effective control over $M \sim
  10^{11}$~ton of atmosphere  may be achieved either from the
  earth (see Fly's Eye \cite{FlysEye} and HiRes \cite{HiRes}) or
  from satellites (Airwatch \cite{Airwatch} etc);
  \item using such a 'neutrino' detector, as \emph{Pierre Auger}
  installation\footnote{A hybrid detector in Argentina will
  consist of $1600$ tanks, each filled with $12$~m$^3$ of water.
  Stations will be distributed in a grid with $1.5$~km spacing. b)
  Four \emph{Fly's Eye} type fluorescence detectors, controlling
  $\sim 3000$~km$^2$ of the site.} \cite{Auger}. Note, that
  \emph{Pierre Auger} telescope is designed for the study of \CRs\
  at $E > 1\times 10^{20}$~eV, high-energy \n\ physics being just
  a byproduct.
\end{itemize}
\subsection{Rate of events} The rate of nucleon-electromagnetic
cascades in a \n\ detector is proportional to convolution product
of $\nu N$-\css\ and $\nu$-fluxes, and also to the number of
scattering centers, i.e.\ mass of detector. A standard requirement
is to have at least $10$ events per year, but the only way to gain
higher statistics is to increase the installation mass. But huge
arrays are very expensive. However, the higher are \nuN-\cs, the
higher is the expected rate of events, the better results may be
obtained.
\subsection{Background}
A severe problem for \HENA\ is high \CR\ background. One is to
shield a detector either burying it deep underground (underwater,
underice) or selecting just down-up going events, using the Earth
as a shield. In any case, the atmospheric \n\ background (mostly
from prompt $\nu$'s) is inevitable. Nevertheless, many
$\nu$-source models suggest that extragalactic diffuse $\nu$-flux
dominates at $E \gtrsim 10^{15} \div 10^{17}$~eV.
\subsection{The aims of the paper}
We shall discuss several consequences for \HENA\ of accelerated
growth of \nN\ \css\ at extremely high energies. We argue that
such acceleration may be induced by non-perturbative hard pomeron
with intercept $\sim 1.4$, which was proposed and successfully
exploited in series of recent papers by A. Donnachie and P. V.
Landshoff \cite{DL}. We study its influence on the rate of
cascades in a $\nu$-detector and on evolution of \n\ spectra
during their pass through large column depths of matter.
\section{Cross-sections}
Neutrino-nucleon scatterings proceed via $CC$ and $NC$
interactions:
\begin{eqnarray}%
\nu_l(\bar{\nu}_l) + N \rightarrow & l^\mp + X,%
\quad \quad & (CC) \label{CC} \\%
\nu_l(\bar{\nu}_l) + N  \rightarrow & \nu_l(\bar{\nu}_l) + X, %
\quad \quad &(NC) \label{NC} %
\end{eqnarray}%
where $l=e,\mu,\tau$. High energy behavior of \nuN-\css\ is yet
unknown. All attempts to estimate them at $E \gtrsim 1$~TeV are
mostly based on different extrapolations of nucleon
\emph{Structure Functions (SFs)} $F_2^{\nu N}(x,Q^2)$, $F_3^{\nu
N}(x,Q^2)$ to $x \lesssim 1 \times 10^{-5}$ (see e.g.\ Ref.s
\cite{BGZR,Gandhi,FMR,KMS}). In papers \cite{GYNPQS,GY} a new
parameterization of these \SFs\ has been derived. It was grounded
on a successful description of $F_2^{ep}(x,Q^2)$ by A. Donnachie
and P. V. Landshoff \cite{DL}, which claimed that small-$x$
$ep$-scattering data of \emph{HERA} \cite{HERA} may be
successfully explained with the help of a simple combination of
several Regge theory inspired leading pole terms. The most
important were so-called \emph{'soft'} (with intercept $\sim
1.08$) and \emph{'hard'} (with intercept $\sim 1.4$, hence very
similar to the perturbative \emph{BFKL} one \cite{BFKL}) pomerons;
first prevails at small $Q^2$, while the latter dominates at large
$Q^2$. Moreover, \DL\ argued that perturbative \QCD\ (\pQCD) fails
at small $x < 10^{-5}$ and even that its validity at $x \sim
10^{-4}\div 10^{-5}$ is a pure fluke. However, the \DL\ approach
is also just a model, neglecting, in particular, all non-leading
poles and cuts in the complex angular momentum $l$-plane. This
results in violation of unitarity and unlimited growth of \css\ at
$E_\nu \rightarrow \infty$.

Using generalization of \DL\ $F_2^{ep}(x,Q^2)$ description to
\nuN-scattering case, \SFs\ $F_2^{\nu N}(x,Q^2)$ and $F_3^{\nu
N}(x,Q^2)$, presumably valid in the whole range of kinematic
variables $0 \leq x \leq 1$ and $0 \leq Q^2 \leq \infty$, have
been derived in Ref.\ \cite{GY}. At $x \gtrsim 10^{-5}$ they were
determined by \pQCD\ parameterization \emph{CTEQ5} \cite{CTEQ},
while in the small-$x$ region they were driven by the analogous
Regge theory inspired description. A special interpolation
procedure have allowed to meet smoothly these different, both over
$x$ and $Q^2$, descriptions of \SFs\ at low and high $x$. In
Ref.~\cite{GYNPQS} these new \SFs\ were denoted by \emph{DL+CTEQ5}
indicating that they had their origin in the interpolation between
\DL\ and \pQCD\ descriptions.

In parallel there were considered in Ref.~\cite{GY} \nuN\ \SFs,
which had been derived by a simple extrapolation of \pQCD\ \SFs\
from $x \ge 1\times 10^{-5}$ to the small-$x$ region. These were
given by formula:
\begin{equation}
\label{log} F_i^{\nu N,Log+CTEQ5}(x<x_{min},Q^2) = F_i^{\nu
N,CTEQ5}(x_{min},Q^2)\left(\frac{x}{x_{min}}\right)^{\beta_i(Q^2)}, %
\end{equation}
\begin{equation}
\label{beta}
\beta_i(Q^2) = \left. \frac{\partial \ln F_i^{\nu N, CTEQ5}(x,Q^2)}%
{\partial \ln x} \right|_{x=x_{min}} ; \;\;\; %
x_{min} = 1\times 10^{-5},
\end{equation}%
$i =2,3$. These \SFs\ smoothly shot to the low-$x$ region from the
\emph{CTEQ5} defined high-$x$ one \cite{BGZR,GYNPQS,GY}. Starting
values of functions and of their logarithmic derivatives over $\ln
x$ had been taken at the $x=x_{min}$ boundary of \emph{CTEQ5}.
This parameterization was designated by \emph{Log+CTEQ5}.
Actually, these \SFs\ are very close to analogous of Ref.\
\cite{Gandhi}.

\begin{figure}[h!]
\begin{minipage}[t]{.47\linewidth}
\includegraphics[width=\linewidth]{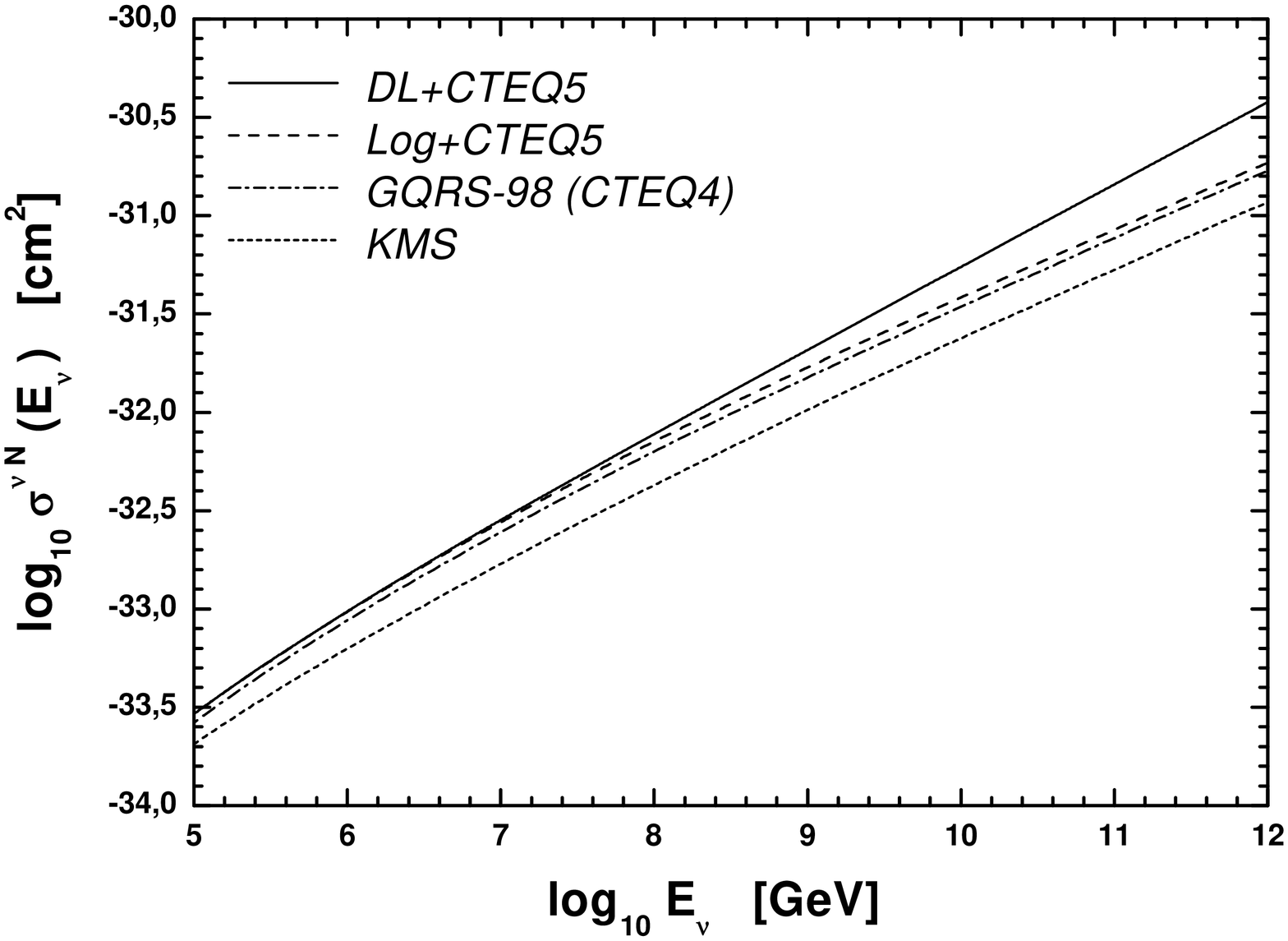}
\caption{Total \emph{(CC+NC)} $\nu N$-\css. Labelling of the
curves is described in the text.} \label{fig:crsscn}
\end{minipage}\hfill
\begin{minipage}[t]{.47\linewidth}
\includegraphics[width=\linewidth]{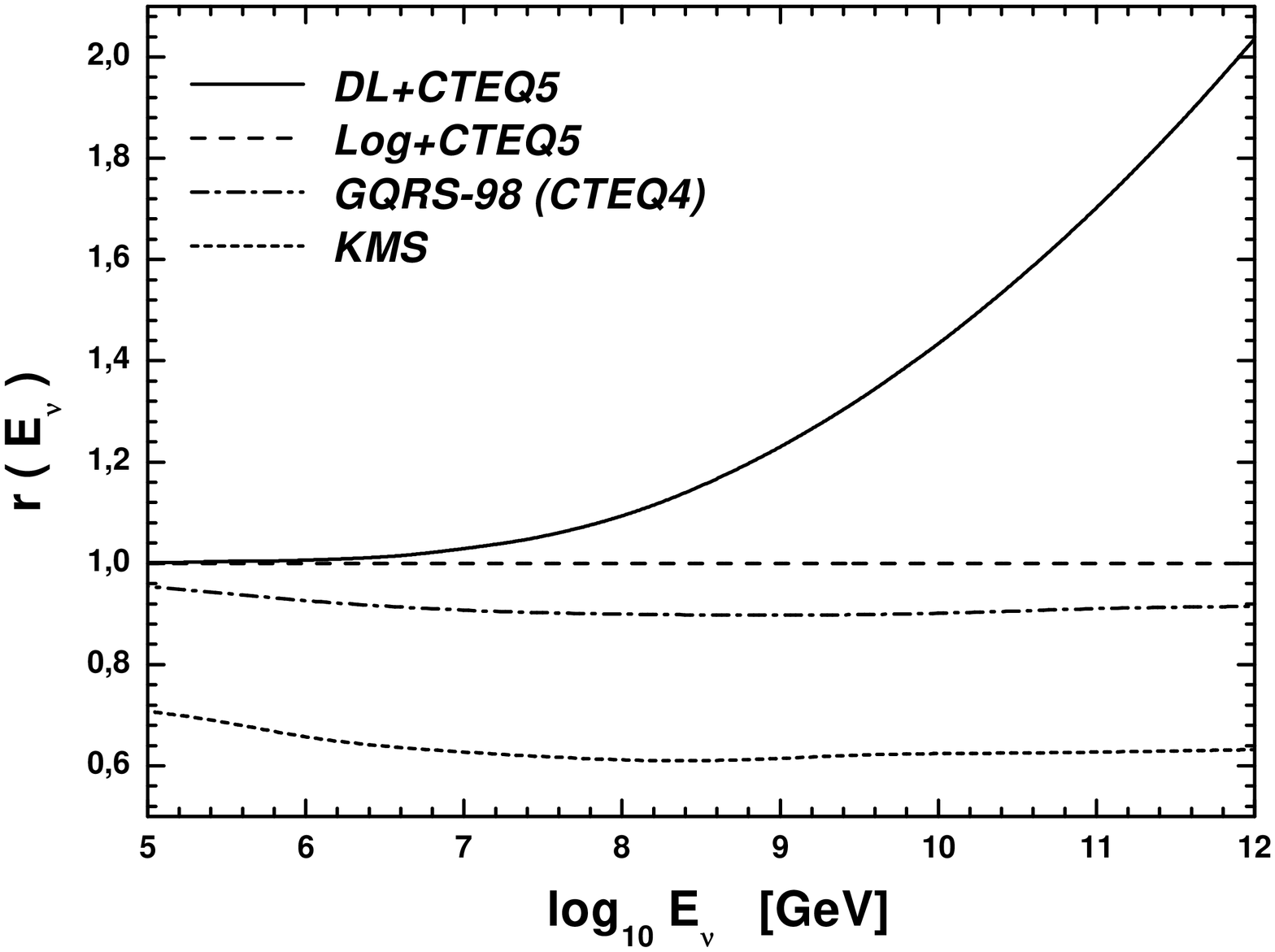}
\caption{ $r(E_\nu)$ are ratios between \css\ shown in
Fig.~\ref{fig:crsscn} and the \emph{Log}+\emph{CTEQ5} \cs.}
\label{fig:rsigma}
\end{minipage}
\end{figure}

Denoted as \emph{DL+CTEQ5}, total  $CC+NC$ \nuN-\css\ are shown in
Fig.~\ref{fig:crsscn}. For comparison there are also plotted the
corresponding cross-sections, which were obtained
\begin{description}
 \item[a)] in the framework of simple $Log + \emph{CTEQ5}$
 extrapolation (\ref{log});
 \item[b)] with the help of \emph{CTEQ4} parameterization by \emph{Gandhi et al.}
 Ref.~\cite{Gandhi}, denoted as \emph{GQRS-98 (CTEQ4)};
 \item[c)] within a united \emph{BFKL/DGLAP} approach by \emph{Kwiecinski,
 Martin and Stasto} \cite{KMS},
labelled as \emph{KMS}.
\end{description}

Due to hard pomeron, our model predicts the most rapid growth of
\css\ at high energies. The differences between cited calculations
become especially clear in Fig.~\ref{fig:rsigma}, where each \cs\
is divided by the \cs\ of \emph{Log}+\emph{CTEQ5}.
\section{Rates of cascades in $\nu$-detectors}
In this section we shall check the manifestations of pomerons in
the rates of nuclear-electromagnetic cascades in deep
underwater/underice detectors. To simplify the problem, we shall
deal with an incident $\nu$-flux energy spectrum, approximated by
a simple power-law decreasing function,
\begin{equation}\label{spectra}
F_\nu(E_\nu,\gamma) = A \times E_\nu^{-(\gamma+1)},
\end{equation}
where $\gamma$ is the index of an integral neutrino spectrum,
$F_\nu(>E,\gamma)$\footnote{ Note that for power-law spectra
$\gamma \times F_\nu(>E,\gamma) = E \times F_\nu(E,\gamma)$.}, and
$A$ is the normalization coefficient. Most source models commonly
assume $1.1 \leq \gamma \leq 2.1$. The real generation
$\nu$-spectrum should be cut-off at high energies of course,
though in top-down models this may occur only at very large
energy. Moreover, as it will be shown later, the high-energy part
of \n\ spectrum suffers from attenuation on passage through large
column depths of matter. Nevertheless, for the purposes of present
consideration all these details may be temporarily neglected.

So, in the case of power-law decreasing spectra (\ref{spectra}),
differential and integral rates of cascade production in a
detector may be calculated with the help of so-called
demensionless differential,
\begin{equation}\label{dhadrmom}
Z_h(E_h,\gamma) = \int \limits_0^1 dy y^\gamma \frac{d\sigma_{\nu
N}(E_h/y,y)}{\sigma_0 dy},
\end{equation}
and integral,
\begin{equation}\label{ihadrmom}
Y_h(E_h,\gamma) = \gamma \int \limits_0^1 du u^{\gamma-1}
Z_h\left(\frac{E_h}{u},\gamma\right),
\end{equation}
hadron moments \cite{NuNBG,Numom}. Here $E_h$ is the energy of a
hadron-electromagnetic cascade, $y = E_h/E_\nu$, and
$\sigma_0=G_F^2 m_W^2/\pi$ is the normalization cross-section (for
$m_W = 81$~GeV $\sigma_0 = 1.09 \times 10^{-34}$~cm$^2$). These
rates in the case of $CC$-scattering (\ref{CC}) are
\begin{equation}\label{drateofcasc}
  dN_h(E_h)/dt = Z_h^{CC}(E_h,\gamma) N_N \sigma_0 \Omega F_\nu(E_h),
\end{equation}
\begin{equation}\label{irateofcasc}
  dN_h(>E_h)/dt = Y_h^{CC}(E_h,\gamma) N_N \sigma_0 \Omega F_\nu(>E_h),
\end{equation}
where $N_N$ is the number of nucleons in a detector, $\Omega$ is
the effective solid angle, opened for the neutrino flux. To take
into account the $\bar{\nu} N$-  and $NC$-interaction cases, one
should just substitute the appropriate differential \css\ in the
Eq.~(\ref{dhadrmom}) for the $CC$ \cs.

In the case of electron (anti)neutrino $CC$-scattering (\ref{CC})
the role of differential hadron moment belongs to the
corresponding normalized $CC$-cross-section, $\sigma_{\nu_e
N}^{CC}(E_\nu)/\sigma_0$; since here $E_h=E_\nu$, the $\nu_e$-flux
in (\ref{drateofcasc},\ref{irateofcasc}) is to be taken at the
energy of incident neutrino.

\begin{figure}[h!]
\begin{minipage}[t]{.47\linewidth}
\includegraphics[width=\linewidth]{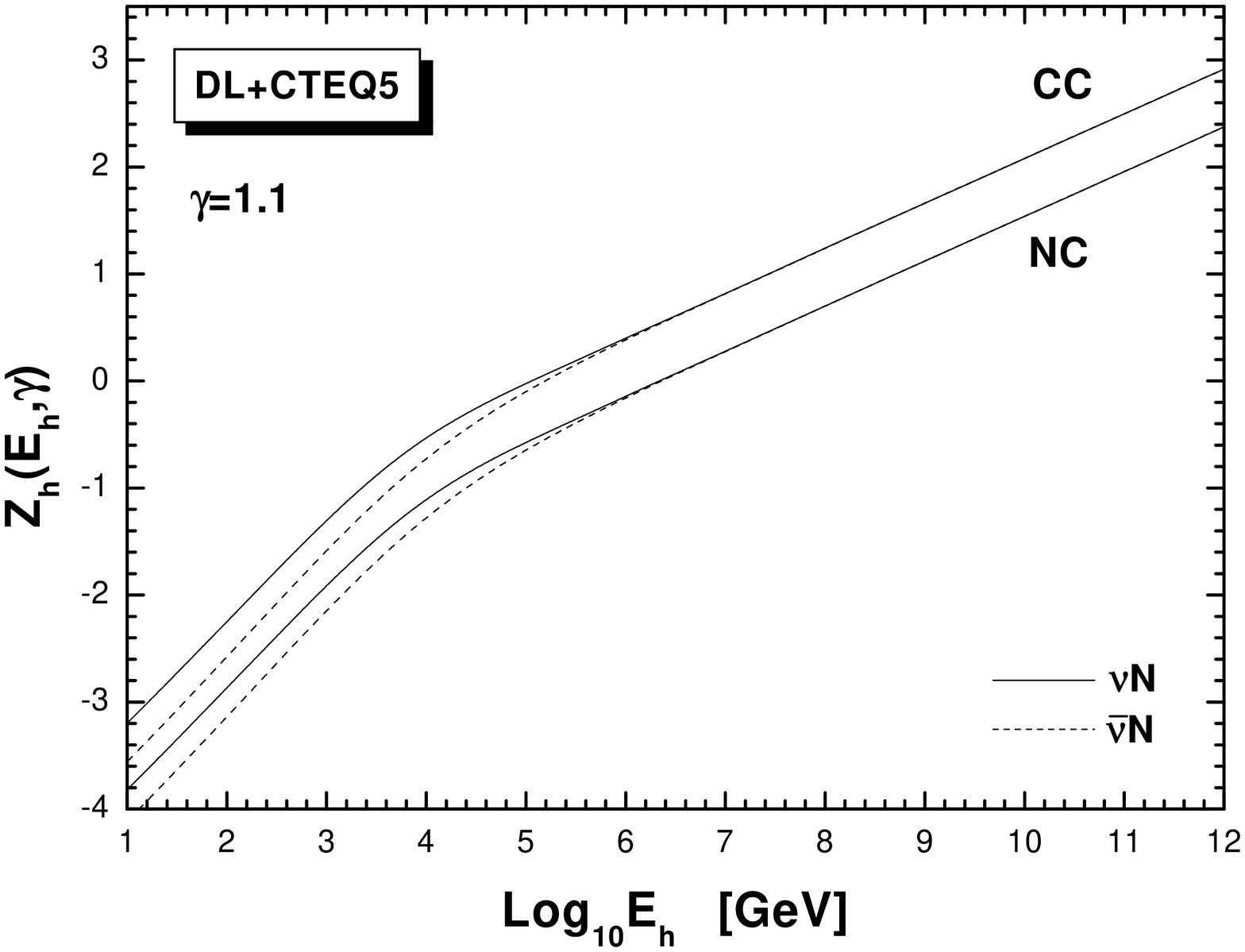}
\caption{Differential $CC$ and $NC$ hadron moments for
$\gamma=1.1$.} \label{fig:zh11}
\end{minipage}\hfill
\begin{minipage}[t]{.47\linewidth}
\includegraphics[width=\linewidth]{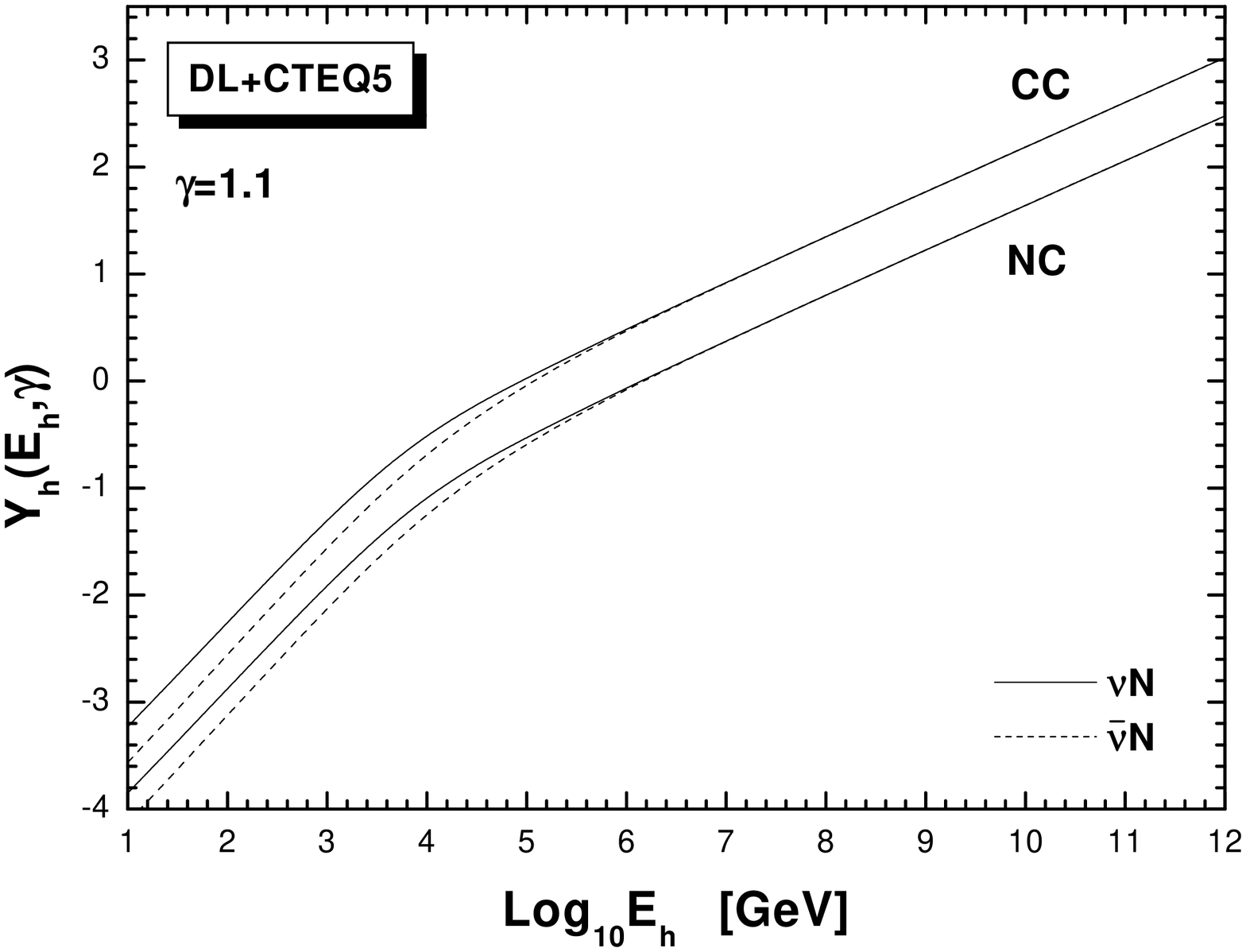}
\caption{Integral $CC$ and $NC$ hadron moments for $\gamma=1.1$.}
\label{fig:yh11}
\end{minipage}
\end{figure}

According to Eq.s~(\ref{dhadrmom},\ref{ihadrmom}), hadron moments
are sensitive to the high-energy parts of $\nu$-spectra. As a
consequence, hard pomeron effects look even more pronounced in
these observables. To illustrate a common trend of hadron moments,
the differential and integral \nuN\ and $\bar{\nu} N$ hadron
moments are plotted in Fig.~\ref{fig:zh11} and
Fig.~\ref{fig:yh11}, respectively, for \emph{DL+CTEQ5}
parameterization and $\gamma = 1.1$, both for $CC$ and $NC$.

A detailed discussion of dependencies of hadron moments on $\gamma$ can be found in
Ref.~\cite{GYNPQS10}. \\

\begin{figure}[h!]
\begin{minipage}[t]{.47\linewidth}
\includegraphics[width=\linewidth]{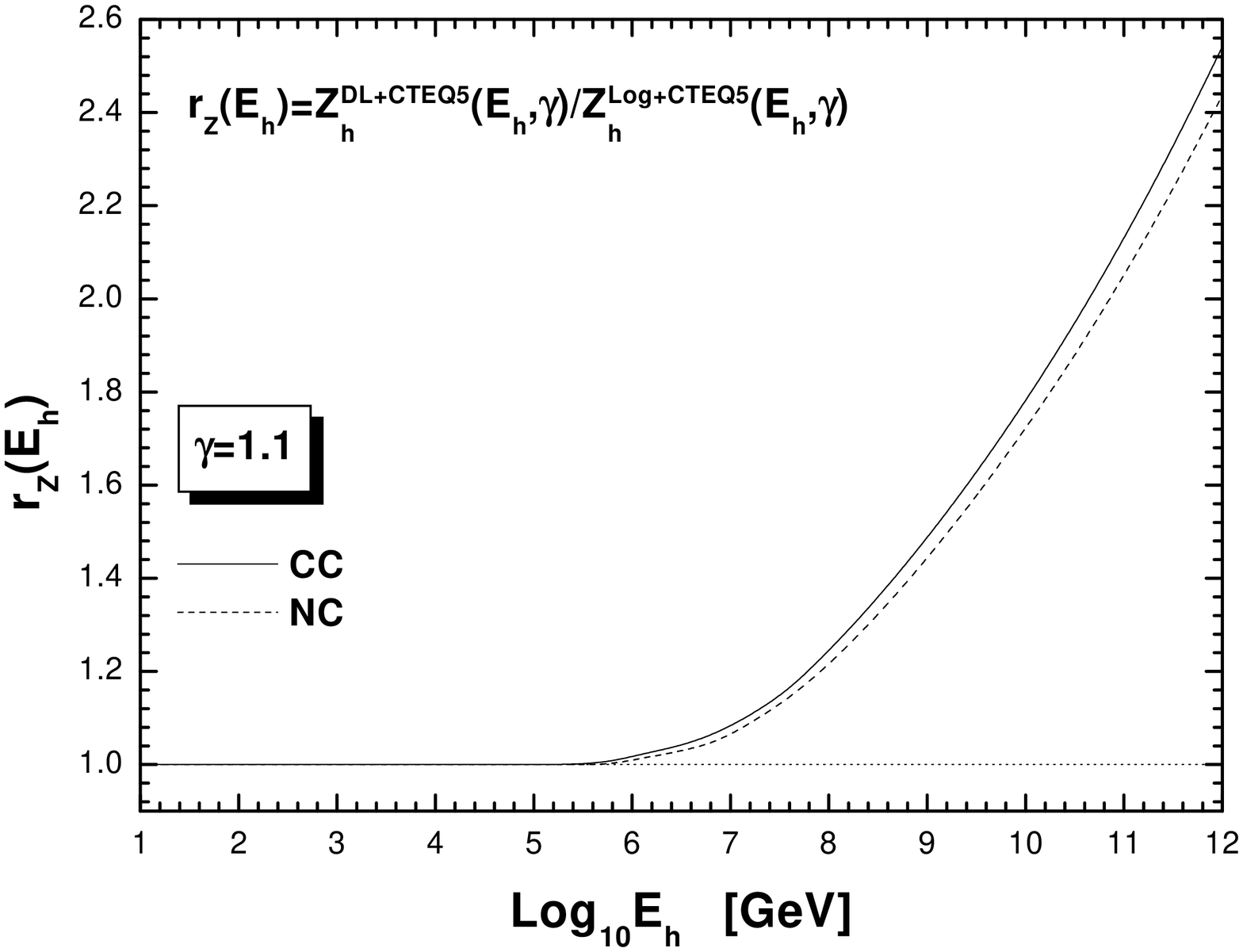}
\caption{Ratios between $CC$ differential hadron moments of
\emph{DL+CTEQ5} and \emph{Log+CTEQ5} parameterizations for
$\gamma=1.1$. } \label{fig:rzDL}
\end{minipage}\hfill
\begin{minipage}[t]{.47\linewidth}
\includegraphics[width=\linewidth]{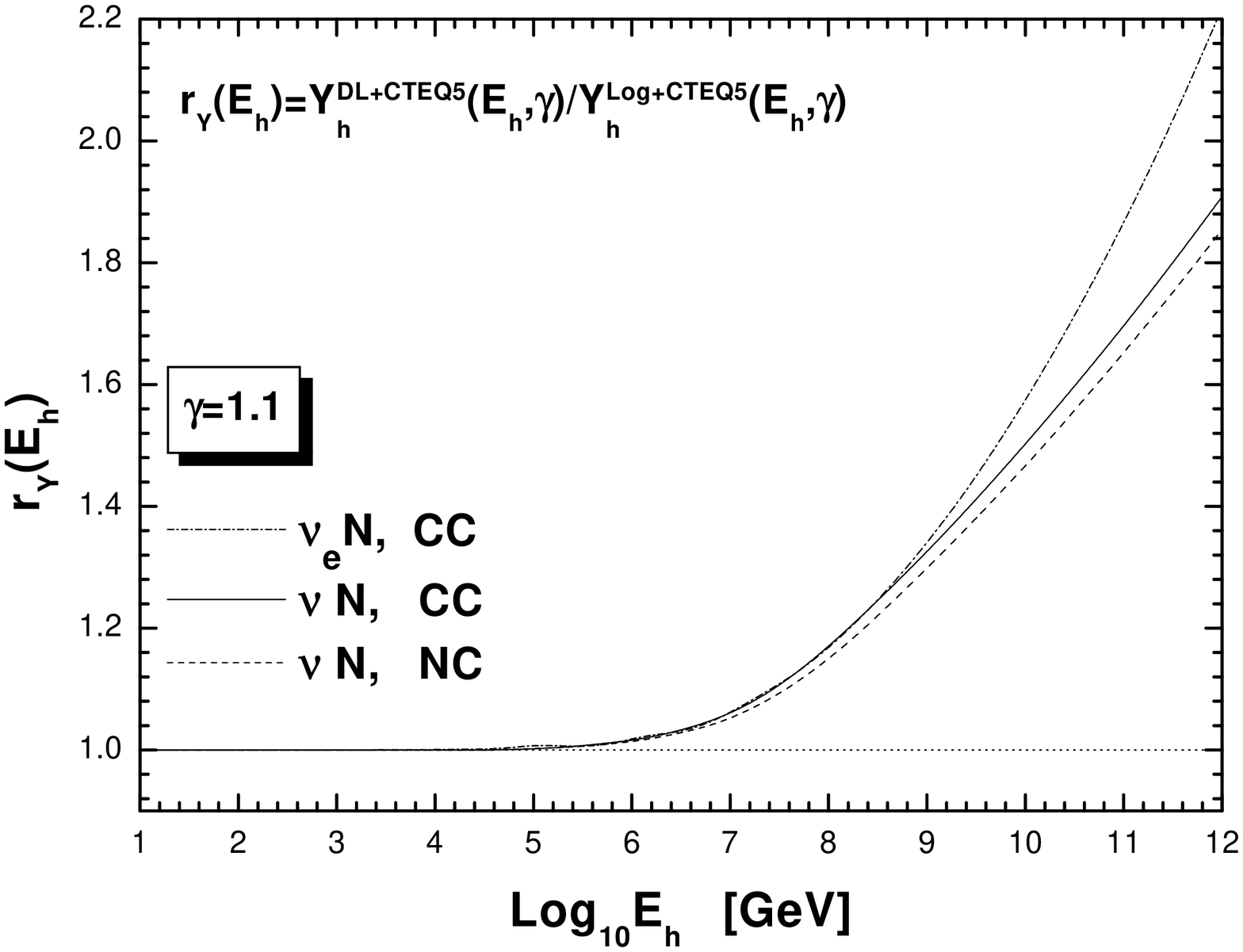}
\caption{Analogous to Fig. \ref{fig:rzDL} ratios of integral
moments. The curve denoted $\nu_e N$ describes the case of
$E_h=E_\nu$.} \label{fig:ryDL}
\end{minipage}
\end{figure}
It is informative to plot ratios between corresponding $CC$ hadron
moments, evaluated within \emph{DL+CTEQ5} and \emph{Log+CTEQ5}
parameterizations, respectively. They are shown in
Fig.s~\ref{fig:rzDL},\ref{fig:ryDL} for $\gamma=1.1$. These
figures corroborate our suggestion, that hadron moments are more
sensitive to the non-perturbative pomeron effects. The strongest
effect in integral moments takes place for $\nu_e(\bar{\nu}_e) N$
$CC$-scattering, when the whole energy of the incident \n\ is
transferred into cascade.

\section{Absorption and regeneration of neutrino fluxes in matter}
Neutrinos are known to be weak interacting particles. However,
\HE\ \ns\ should be absorbed in large dense samples of matter,
when $\sigma_{\nu N}(E)\times n \times l/m_N \gtrsim 1$; here $n$
(in $g/cm^3$) is the matter density and $l$ (in $cm$) is the range
of \n\ pass. An appropriate variable for such consideration is
column depth, $\tau= n\times l$ (measured in $g/cm^2$).

Attenuation effects may be described by a shadow factor, which is
a ratio of \n\ flux at column depth $\tau$ and the primary flux
(i.e.\ at $\tau=0$):
\begin{equation}\label{shadfac}
  S(E,\tau) = \frac{F_\nu(E,\tau)}{F_\nu(E,0)}.
\end{equation}
To a first approximation, shadow factor $S(E,\tau) \approx \exp(-
\sigma^{CC+NC}(E) \tau /m_N)$. It provides an estimate that \ns\
with $E>10$~TeV (see Fig.~\ref{fig:crsscn}), passing through the
center of the Earth (column depth of matter in this case is
$\tau_\bigoplus \simeq 1.1\times10^{10} \mbox{g/cm}^{2}$), are
partially absorbed.

On the one hand, the higher is the energy, the stronger is
attenuation. On the other hand, Earth is the natural, but not the
unique example of a large column depth of matter in the Universe.
One can easily imagine different environments with much higher
column depths, where neutrino flux attenuations would be stronger
as well. Hence, attenuation effects should be taken into account
in \HENA.

In this section we emphasize that hard pomeron enhanced \nuN-\css\
bring to much higher attenuations of \n\ fluxes, than it was
previously expected.

The absorption of \ns\ in the Earth have been earlier discussed by
many authors. However, only in paper \cite{BGZR} the regeneration
of \n\ fluxes due to $NC$-scattering (\ref{NC}) has been taken
into account for the first time. To evaluate the effect, a 'thin
target' model have been used in that paper. That model assumed
that average attenuation length in the Earth is comparable with
its diameter. And that was the case for the Earth as a target and
the energy range considered in Ref.~\cite{BGZR}.

Nevertheless, in order to treat correctly the higher energies and
larger column depths, a more sophisticated consideration is
necessary. Ten years later it was proposed by A. Nicolaidis and A.
Taramopoulos in Ref.~\cite{NicTar}. They have derived a trivial
transport equation
\begin{equation}\label{transpeq}
\begin{array}{ll}
  \frac{d}{d\tau}F_\nu(E_\nu,\tau) = & -\sigma^{CC+NC}(E_\nu)\times
   F_\nu(E_\nu,\tau) + \\
  & + \int_0^1 \frac{dy}{1-y}\times \frac{d\sigma^{NC}(E_\nu/(1-y),y)}{dy}
  \times F_\nu(E_\nu/(1-y),\tau)
\end{array}
\end{equation}
and solved it (unfortunately, under some wrong assumptions about
differential cross-sections). First term in the r.h.s.\ of the
Eq.~(\ref{transpeq}) accounts for the absorption, while the second
one describes regeneration via $NC$.

Later these equations have been solved correctly by many authors
with various parameterizations of high-energy \nuN-\css\ (see
e.g.\ Ref.s~\cite{Gandhi,KMS}).

An elegant method for solution of the transport equation
(\ref{transpeq}), along with a detailed discussion of different
\n\ spectra evolution, has been given by V. A. Naumov and L.
Perrone in paper \cite{NaumPer}. In this paper we exploited their
method, applying it to the infinite incident power-law decreasing
\n\ spectra (\ref{spectra}). Calculations were performed for the
two parameterizations of differential \css, viz.\ for $DL+CTEQ5$
and $Log+CTEQ5$.

\begin{figure}[h!]
\begin{minipage}[t]{.44\linewidth}
\includegraphics[width=\linewidth]{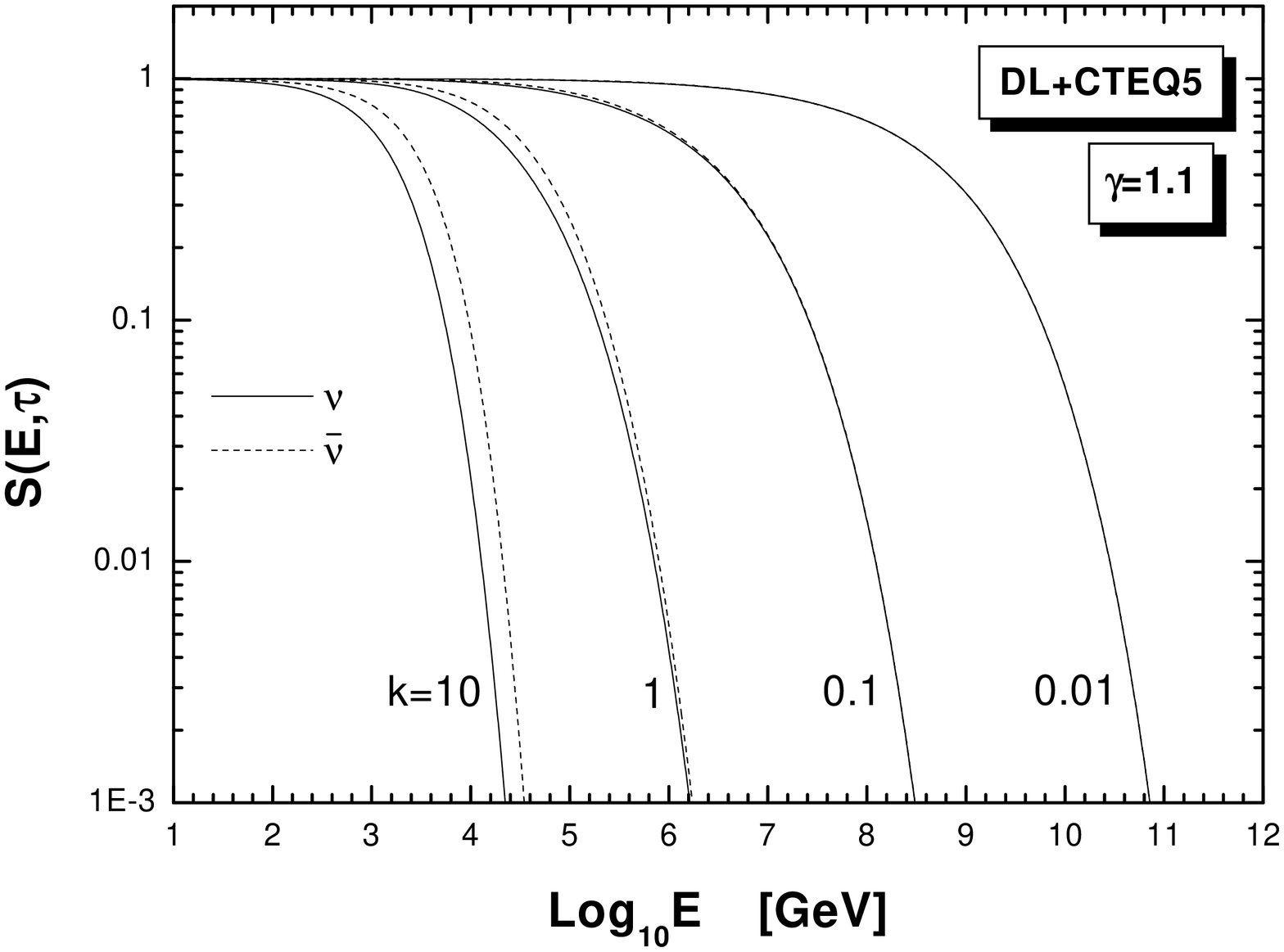}
\caption{Shadow factors for \nuN- and $\bar{\nu}N$-scattering for
several values of column depth, parameterized by the factor
$k=\tau/\tau_\bigoplus$; $\tau_\bigoplus=1.1\times10^{10}
\mbox{g/cm}^{2}$ corresponds to the passage through the center of
the Earth.} \label{fig:shnan}
\end{minipage}\hfill
\begin{minipage}[t]{.44\linewidth}
\includegraphics[width=\linewidth]{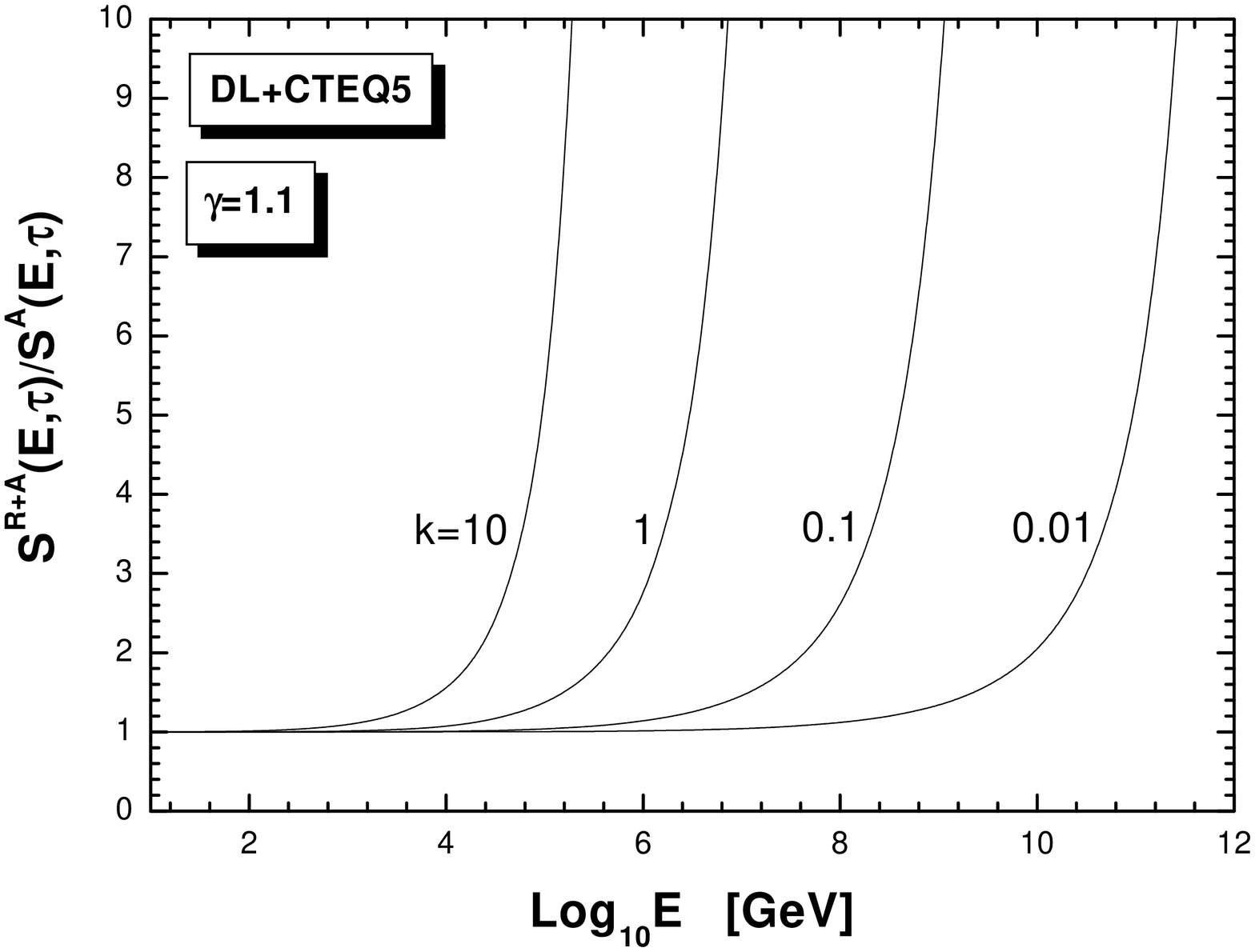}
\caption{Ratios of \nuN-scattering shadow factors calculated with
and without regard for regeneration of $\nu$-fluxes by neutral
currents. Factors $k$ are the same as in Fig.~\ref{fig:shnan}.}
\label{fig:shabsreg}
\end{minipage}
\end{figure}

\begin{figure}[h!]
\begin{minipage}[t]{.47\linewidth}
\includegraphics[width=\linewidth]{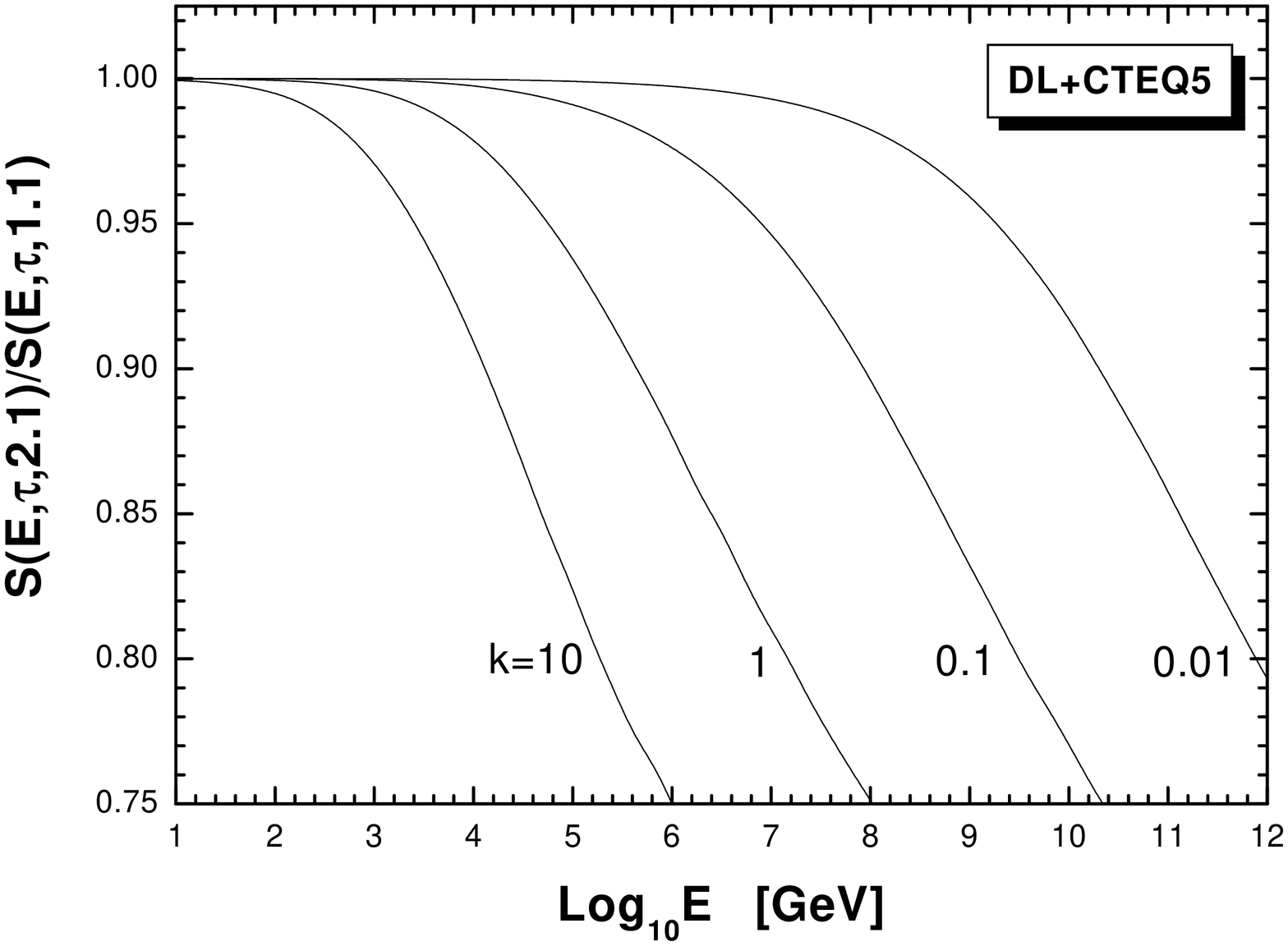}
\caption{Ratios of \nuN\ shadow factors for incident $\nu$-spectra
with indices $\gamma = 2.1$ and $\gamma = 1.1$, respectively.
Factors $k$ are the same as in Fig.~\ref{fig:shnan}.}
\label{fig:sh2111}
\end{minipage}\hfill
\begin{minipage}[t]{.47\linewidth}
\includegraphics[width=\linewidth]{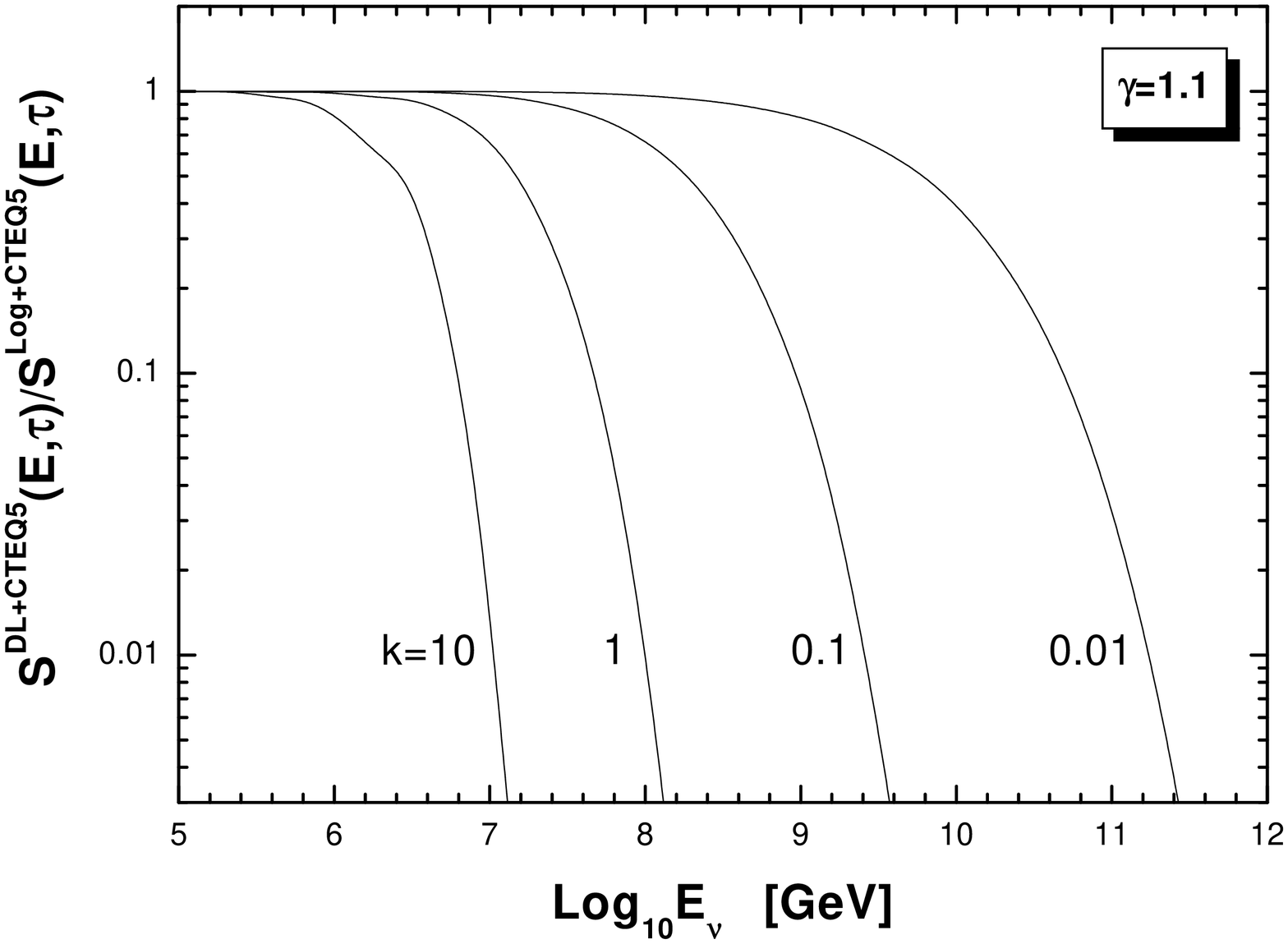}
\caption{Ratios of \nuN\ shadow factors, calculated within
$DL+CTEQ5$ and $Log+CTEQ5$ parameterizations, respectively.
Factors $k$ are the same as in Fig.~\ref{fig:shnan}.}
\label{fig:shDLLog}
\end{minipage}
\end{figure}

Our results are demonstrated in Fig.s~\ref{fig:shnan}-\ref{fig:shDLLog}. \\

In Fig.~\ref{fig:shnan} shadow factors $S(E,\tau)$
(\ref{shadfac}), evaluated with the help of hard pomeron enhanced
parameterization $DL+CTEQ5$, are plotted for \ns\ and anti\ns. All
curves relate to only one value of initial \n\ spectrum index,
$\gamma =1.1$. Evolution is illustrated by successively increasing
values of column depth $\tau$, which are expressed in terms of the
'standard' column depth $\tau_\bigoplus = 1.1 \times10^{10}
\mbox{g/cm}^{2}$.

In Fig.~\ref{fig:shabsreg} the importance of the regeneration term is emphasized. The
ratios of shadow factors with regeneration and those accounting only for absorption
are shown here for the same successively increasing column depths $\tau$ as in
Fig.~\ref{fig:shnan}. \\

The dependencies of shadow factors on $\gamma$ may be understood
from Fig.~\ref{fig:sh2111}. The ratios of these factors are
plotted here for two indices of initial spectra, viz.\
$\gamma=2.1$ and $\gamma=1.1$. Evolution is shown in the same way
as it was done in the previous figures.

And, finally, the ratios of shadow factors calculated within
$DL\-+CTEQ5$ and $Log+CTEQ5$ parameterizations are shown in
Fig.~\ref{fig:shDLLog}. Notations for evolution remain the same as
in the previous pictures. This plot persuades that at high
energies the hard pomeron effects cannot be neglected even for
rather shallow column depths of matter. Pomeron effects, being
raised in shadow factors into exponent, bring to tremendous
changes as compared with previous estimations. These coefficients
are evidently the most sensitive observables in the \HE\ \n\
astrophysics to the hypothetical hard pomeron enhancement of \nuN\
\sfts.

\section*{Conclusions}
In this paper we have studied different manifestations of
hypothetical hard pomeron enhanced \nuN\ structure functions in
the high-energy neutrino astrophysics. We have compared i) \css\
$\sigma_{\nu N}(E)$, ii) rates of hadron-electromagnetic cascades,
parametrized through differential and integral hadron moments,
$Z_h(E_h,\gamma), Y_h(E_h,\gamma)$, and iii) shadow factors
$S(E_\nu,\tau)$, calculated within hard pomeron enhanced
$DL+CTEQ5$ parameterization, with analogous values calculated
within simple extrapolation of structure functions from high-$x$
perturbative \QCD\ description to the low-$x$ region, $Log+CTEQ5$.

We found that these effects, if our assumption are justified, may
change substantially our notion of high energy neutrino
astrophysics. The most pronounced evidence for their presence has
been found in the rapid decrease of shadow factors with energy. It
may be seen even with column depths available at the Earth. We
hope the validity of our suggestions will be checked by future
giant neutrino telescopes.

\section*{Acknowledgments}
This research has been supported in part by the \emph{INTAS} grant
No: 99-1065.

\end{document}